\documentstyle[15lomcon,cite,epsfig]{article}

\begin{document}

\title{FIRST LHC CONSTRAINTS ON ANOMALOUSLY INTERACTING NEW VECTOR BOSONS}
\author{Mikhail Chizhov} 
\address{Centre for Space Research and Technologies, 
University of Sofia, 
Bulgaria}
\author{Vadim Bednyakov \footnote{e-mail: bedny@jinr.ru},
Igor Boyko, Julian Budagov, Mikhail Demichev, \\ and Ivan Yeletskikh}
\address{Dzhelepov Laboratory of Nuclear Problems,\\
\mbox{Joint Institute for Nuclear Research, 141980, Dubna, Russia}}

\maketitle\abstracts{
     It was recently proposed to extend the Standard Model 
     by means of new spin-1 chiral $Z^*$ and $W^{*\pm}$ bosons
     with the internal quantum numbers of the electroweak Higgs doublets. 
     These 
     bosons have unique signatures in transverse momentum,
     angular and pseudorapidity distributions of the final leptons,
     which allow one to distinguish them from other heavy resonances.
     With 40~pb$^{-1}$ of the LHC proton-proton data at the 
     energy 7~TeV, the ATLAS detector was used to search for narrow resonances
     in the invariant mass spectrum 
     of $e^+e^-$ and $\mu^+\mu^-$ final states and high-mass charged states
     decaying to a charged lepton and a neutrino.
     From the search 
     exclusion mass limits of 1.15~TeV$/c^2$ and 1.35~TeV$/c^2$
     were obtained for the chiral neutral $Z^*$ and charged $W^*$ bosons, respectively.
     These are the first direct
     limits on the $W^*$ and $Z^*$ boson production.
}

\section{Chiral boson model} 
      New heavy neutral gauge bosons are predicted in many extensions of 
      Standard Model (SM).
      They are associated with additional U(1)$'$ gauge symmetries and
      are generically called $Z'$ bosons.
      The minimal gauge interactions of these bosons with matter lead to
      the well-known angular distribution of outgoing leptons (the $Z'$ decay product)
      in the dilepton center-of-mass reference frame.
       In addition, another type of spin-1 bosons may exist, which leads to
       a different signature in the angular distribution.
       This follows from the presence of different types of relativistic spin-1 
       fermion currents $\bar{\psi}\gamma^\mu(1\pm\gamma^5)\psi$ and
       $\partial_\nu[\bar{\psi}\sigma^{\mu\nu}(1\pm\gamma^5)\psi]$, which
       can couple to the corresponding bosons.
      The mesons assigned to the tensor quark states are some types of
      ``excited'' states as far as the only orbital angular momentum with
      $L=1$ contributes to the total angular momentum, while the total spin
      of the system is zero.
      This property manifests itself in their derivative couplings 
      to matter and a different chiral structure of the anomalous 
      interactions in comparison with the minimal gauge ones.

      Let us assume that the electroweak gauge sector of the
      SM is extended by a doublet of new spin-1 {\em chiral\/}
      bosons $\mbox{\boldmath$W$}^*_\mu$ with the internal quantum numbers
      of the SM Higgs boson.
      There are at least three different classes of theories,
      all motivated by the Hierarchy problem, which predict
      new vector weak doublets with masses not far from the electro-weak scale 
\cite{Chizhov:2009fc}.
       It is possible to point out
      several model-independent and unique signatures which allow one
      to identify production of such bosons at the hadron colliders
\cite{Chizhov:2008tp}.

      Since the tensor current mixes the left-handed and right-handed
      fermions, which in the SM are assigned to different
      representations, the gauge doublet should have only anomalous
      interactions
\begin{equation} \label{master}
    {\cal L}^*\!=\!\frac{g}{M}\!\left(
                     \partial_\mu W^{*-}_\nu  \;
                     \partial_\mu \overline{W}^{*0}_\nu
                 \right) 
                 \overline{D_R}\;\sigma^{\mu\nu}\left(\hspace{-0.2cm}
                                                        \begin{array}{c}
                                                          U_L \\
                                                          D_L
                                                        \end{array}\hspace{-0.2cm}
                                                      \right)
\!+\!
    \frac{g}{M}\left(
    \overline{U_L}\; \overline{D_L}
                        \right)
                        \sigma^{\mu\nu}D_R \! 
                        \left(\hspace{-0.2cm}
                          \begin{array}{c}
                            \partial_\mu W^{*+}_\nu \\
                            \partial_\mu W^{*0}_\nu \\
                          \end{array}\!\!\!\!\right),
\end{equation}
       where $M$ is the boson mass, $g$ is the coupling constant of the
       SU(2)$_{\rm W}$ weak gauge group, and $U$ and $D$ generically denote
       up-type and down-type leptons and quarks.
        These bosons, coupled to the tensor quark currents, can be considered
	as {\em excited\/} states.
      For comparison we will consider topologically analogous gauge
      interactions of the $Z'$ boson
\begin{equation}\label{Z'ed}
{\cal L}'_{NC}=\frac{g}{2}
    \left(\bar{\ell}\gamma^{\mu}\ell+\bar{d}\gamma^{\mu}d\right)Z'_\mu
\end{equation}
      with the same mass $M$.
      The coupling constants are chosen in such a way that all fermionic 
      decay widths in the Born approximation of the both neutral
      bosons are identical.
      It means that their total production
      cross sections at the hadron colliders are nearly equal up to
      next-to-leading order corrections.
      Their total fermionic decay width
      $\Gamma=\frac{g^2}{4\pi}M\approx 0.034M $
     is sufficiently narrow and they can be identified as resonances
     in the Drell--Yan process.

    Up to now, any excess in the yield of the Drell--Yan process with
    high-energy invariant mass of the lepton pairs remains the clearest
    indication of possible production of a new heavy neutral boson 
    at the hadron colliders.
     The peaks in the dilepton invariant mass distributions originate from the
     Breit--Wigner propagator form, which is {\em the same}\/
     for both the gauge and chiral {\em neutral}\/
     bosons in the Born approximation.
     Concerning discovery of the {\em charged}\/
     heavy boson at the hadron colliders one believes that the cleanest method
     is detection of its subsequent leptonic
     decay into an isolated high transverse-momentum charged lepton.
     In this case the heavy new boson can be observed
     through the Jacobian peak in the transverse 
     $p^{}_{\rm T}$ (or $m^{}_{\rm T}$) distribution.
     It has become proverbial 
     that the Jacobian peak is an inevitable 
     characteristic of any two-body decay.
     However, it is not the case for decays of the new chiral bosons
\cite{Chizhov:2006nw}.
      It has been found in
\cite{Chizhov:2000vt} that tensor interactions lead to a new angular 
      distribution of the outgoing fermions
\begin{equation}
\label{GLR} 
    \frac{{\rm d} \sigma(q\bar{q}\to Z^*\!/W^*\to f\bar{f})}
     {{\rm d} \cos\theta} \propto \cos^2\theta,
\end{equation}
      in comparison with the well-known vector interaction result
\begin{equation}
\label{GLL}
    \frac{{\rm d} \sigma(q\bar{q}\to Z'\!/W'\to f\bar{f})}
    {{\rm d} \cos\theta} \propto     1+\cos^2\theta \, .
\end{equation}
     The absence of the constant term in the first case results
     in very new experimental signatures
\cite{Chizhov:2006nw}.
     The angular distribution for vector interactions
(\ref{GLL})
     includes a nonzero constant term, which leads to the kinematical
     singularity in the $p^{}_{\rm T}$ distribution of the final fermion.
     This singularity is transformed into a well-known Jacobian peak due
     to a finite width of the resonance.
     In contrast, the pole in     the decay distribution 
(\ref{GLR})
      of the $Z^*/W^*$ bosons is canceled out and
     the fermion transverse momentum $p_{\rm T}$ distribution
     even reaches zero at the
     kinematical endpoint $p_{\rm T}=M/2$.
     A crucial difference between the neutral chiral bosons
     and other resonances should come from the analysis of the angular
     distribution of the final-state leptons with respect to the boost
     direction of the heavy boson in the rest frame of the latter (the
     Collins--Soper frame
\cite{Collins:1977iv}).
       Instead of a smoother angular distribution for the gauge
       interactions 
       a peculiar ``swallowtail'' shape of the chiral boson
       distribution 
       occurs with a dip at $\cos\theta^*_{\rm CS}=0$.
       Neither scalars nor
       other particles possess such a type of angular behavior
(see also 
\cite{Chizhov:2011mb}).

\section{The first experimental constraints on the chiral bosons}
\def\met{$ {E^{\rm miss}_{\rm T }}$}
\def\pt{${p_{\rm T}}$}

\begin{figure}[!htb]
\hspace*{-0.3cm}
\includegraphics[width=0.5\columnwidth]{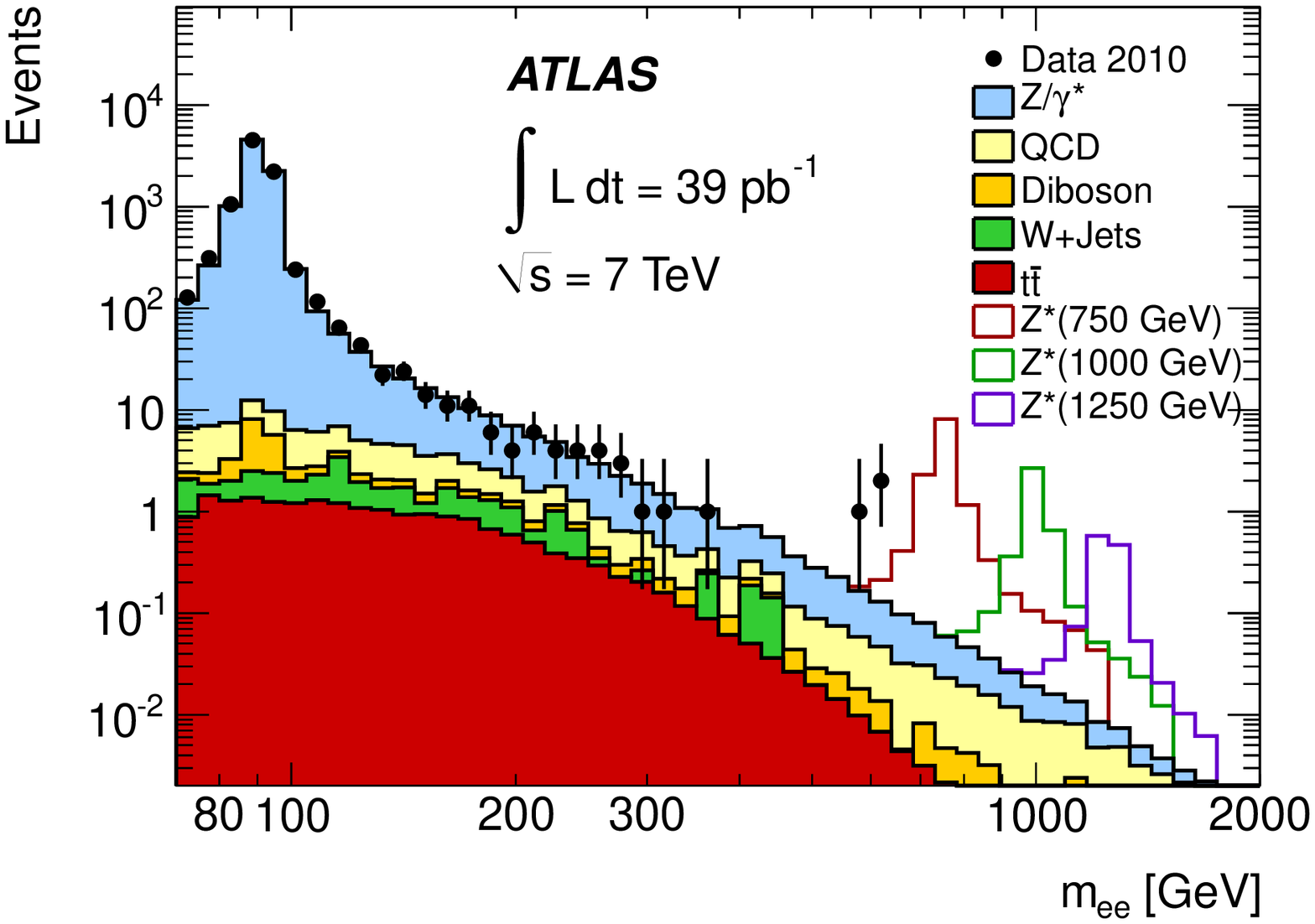}
\includegraphics[width=0.5\columnwidth]{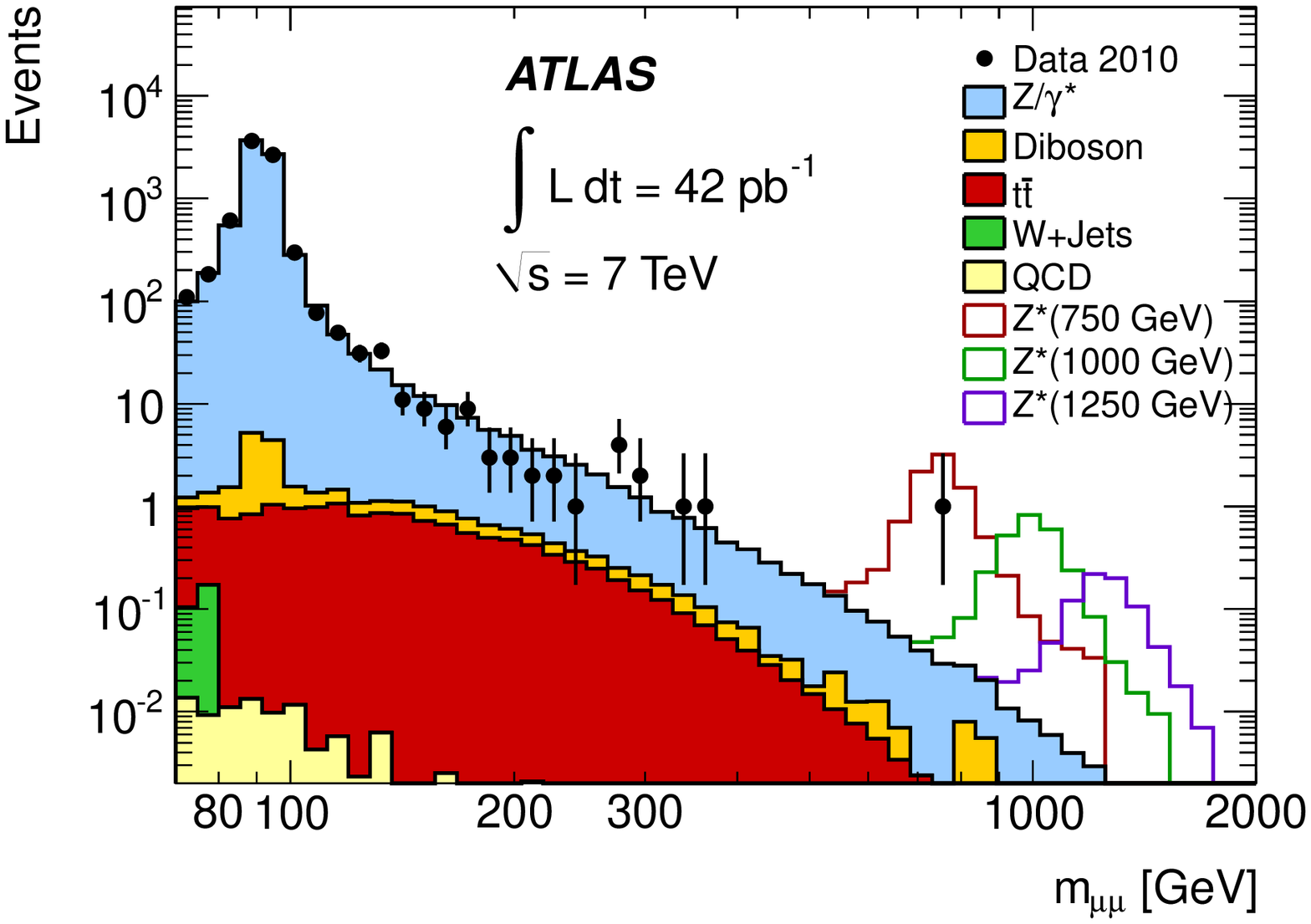}
\vspace*{-0.3cm}
  \caption{Dielectron (left) and dimuon (right)
          invariant mass distribution, 
          compared to the stacked sum of all expected backgrounds and
          with three example $Z^*$ signals overlaid.
}
   \label{fig:ZstarPlots}
\end{figure}

      The first direct experimental search for the excited chiral vector bosons
      was performed by the ATLAS collaboration
\cite{Aad:2008zzm,Collaboration:2010knc} in 2010.
      At the LHC energy of 7 TeV with the 
      integral luminosity around 40~pb$^{-1}$
      the ATLAS detector was used for searching for narrow resonances
      in the invariant mass spectrum above 110~GeV$/c^2$
      of $e^+e^-$ and $\mu^+\mu^-$ final states
\cite{Aad:2011xp}.
      The main physical results relevant to our discussion are presented in 
Fig.~\ref{fig:ZstarPlots}. 
      It is seen that both the dielectron and dimuon invariant 
      mass distributions are well described by the prediction 
      from SM processes.
      Nevertheless, these distributions were 
      for the first time used to
      obtain a lower direct mass limit of 1.152~TeV$/c^2$
      for the neutral chiral $Z^*$ boson. 
       This is the first direct mass limit on this particle.
       The $Z^*$ limits are about 100--200 GeV/$c^2$ more stringent than
       the corresponding limits on all considered $Z'$ bosons.

\def\wps{$ {W'/W^*}$}

\begin{figure*}[!h]
\vspace*{-0.3cm}
  \centering
  \includegraphics[width=0.535\textwidth]{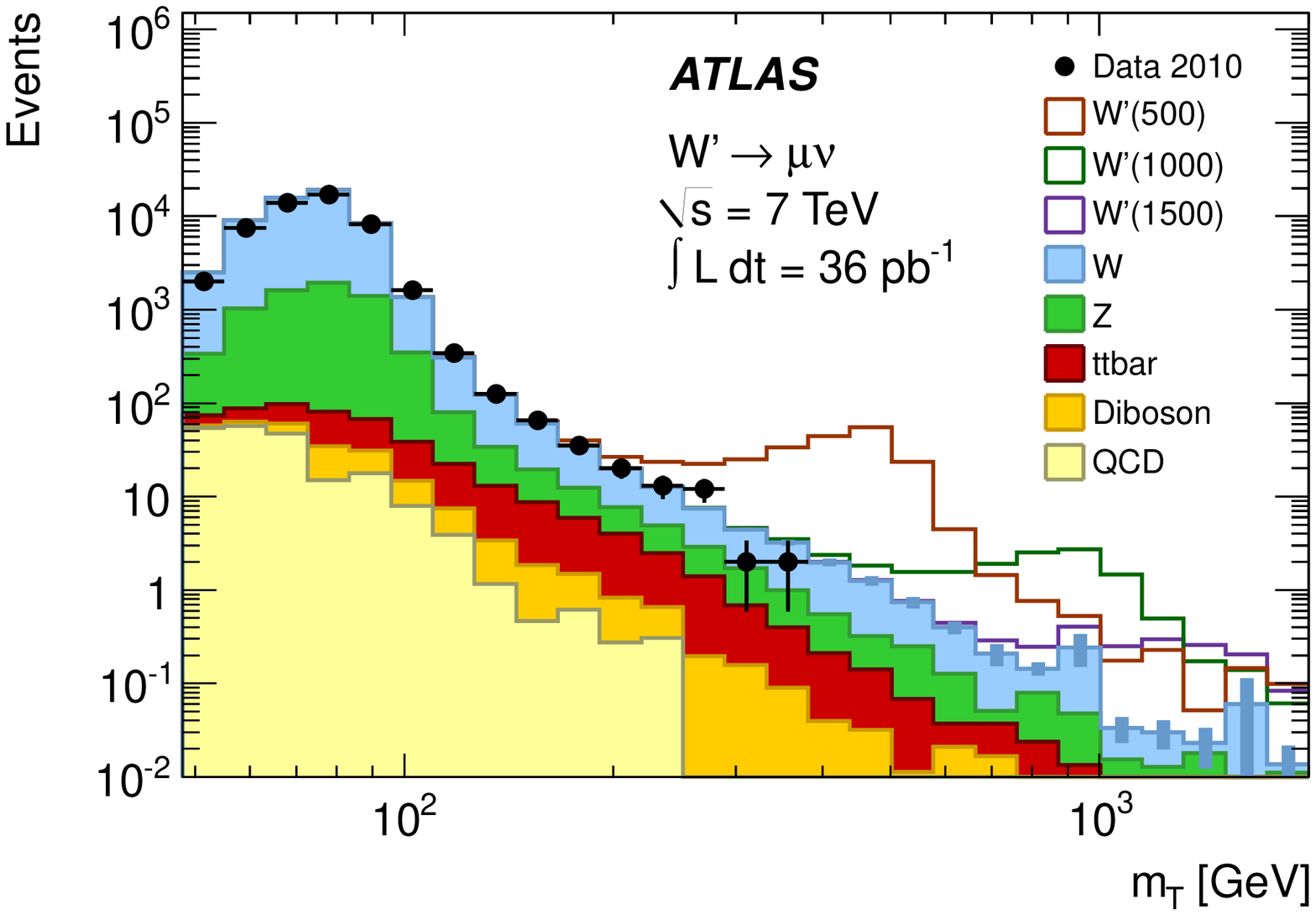}
  \includegraphics[width=0.455\textwidth]{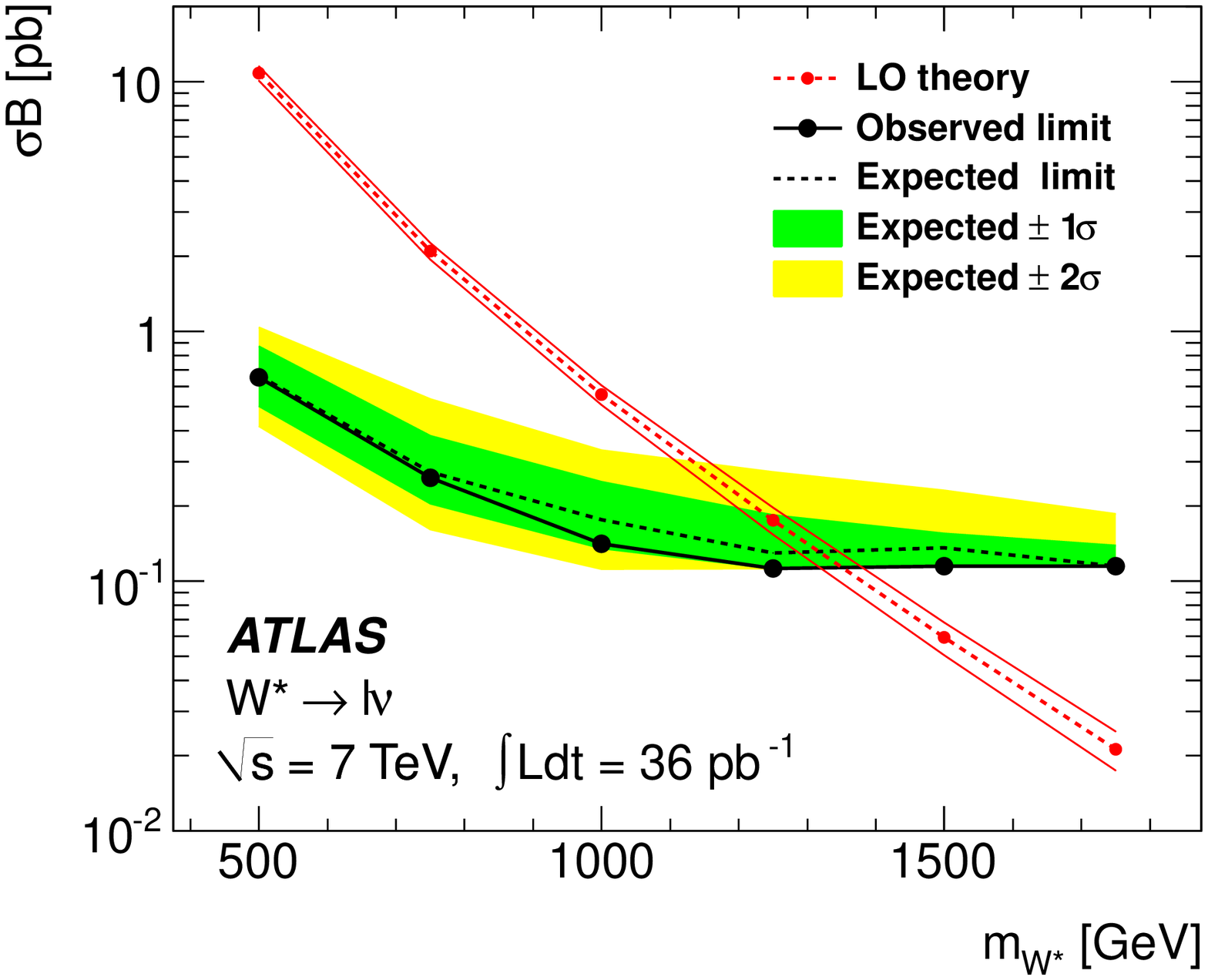}
\vspace*{-0.8cm}
  \caption{
    Spectra of ${m_{\rm T}}$ for the muon decay channel (left panel).
    Limits at 95\% C.L. for  $W^*$  production in the combination of 
    both lepton decay channels (right panel).
    The solid lines show the observed limits with all uncertainties.
From \cite{Aad:2011fe}.
  \label{fig:final_mt}
  }
\vspace*{-0.3cm}
\end{figure*}

      Furthermore, the ATLAS collaboration 
      searched for high-mass states,
      such as heavy charged gauge bosons, 
      decaying to a charged lepton and a neutrino
\cite{Aad:2011fe}.
      The search for heavy charged resonances inclusively produced at the LHC
      looks more complicated than the search for neutral states
      due to the absence of the second decay particle --- the undetectable neutrino.
      In this case
      the kinematic variable used to identify the \wps\ is the transverse mass
$m_{\rm T} = \sqrt{ 2 p_{\rm T} E^{\rm miss}_{\rm T} (1 - \cos \phi_{l\nu})}$.
      Here \pt\ is the lepton transverse momentum,
      \met\ is the magnitude of the missing transverse momentum, 
      and $\phi_{l\nu}$ is the angle between the \pt\ and
      missing ${E_{\rm T}}$  vectors.
      The main physical results 
     relevant to our consideration are given in
Fig.~\ref{fig:final_mt}.
    The agreement between the data and the expected background is rather good.
    The lower mass limits expected and obtained
    from these measurements are depicted
    in the right panel of
Fig.~\ref{fig:final_mt}. 
    The intersection between the central theoretical prediction and the observed limits
    provides the 95\% C.L. lower limit on the mass.
    It was found that the charged chiral $W^*$ 
    boson     was excluded for masses below 1.350~TeV$/c^2$.
    These are the first direct limits on the $W^*$ boson production.

\vspace*{-0.2cm}\section*{References} 

\providecommand{\href}[2]{#2}\begingroup\raggedright\endgroup

\end{document}